\documentclass[journal,10pt]{IEEEtran}

\usepackage{graphicx}
\usepackage{amsfonts}
\usepackage{epstopdf}
\usepackage{amsmath}
\usepackage{algorithm}
\usepackage{caption}

\usepackage{algpseudocode}

\begin{document}

\title{Evolution of Vehicle Network on a Highway}

\author{
Gleb Dubosarskii,~\IEEEmembership{Student Member,~IEEE}, Serguei Primak, \IEEEmembership{Member,~IEEE},\\ Xianbin Wang, \IEEEmembership{Fellow,~IEEE}
\thanks{G. Dubosarskii, S. Primak and X. Wang are with the Department of Electrical and Computer Engineering, Western University,
	London, Ontario, Canada, N6A 5B9 (e-mail: gdubosar@uwo.ca; slprimak@uwo.ca; \mbox{{xianbin.wang@uwo.ca}}).}%
\thanks{Manuscript received November 30, 2018; revised April 1, 2019.}}

\markboth{IEEE Transactions on Vehicular Technology,~Vol.~XX, No.~XX, XXX~2015}
{}
%{Shell \MakeLowercase{\textit{et al.}}: Bare Demo of IEEEtran.cls for Journals}

\maketitle

\begin{abstract}
One of the challenges related to the investigation of vehicular networks is associated with predicting a network state regarding both short-term and long-term network evolutionary changes. This paper analyzes a case in which vehicles are located on a straight road, and the connectivity state between two consecutive cars is determined by the Markov chain model with two states. The transition probabilities of the considered model are explicitly expressed in terms of known parameters of the network using the Wang-Moayery model. Within the presented model, the network evolution is described in terms of determinative parameters, such as average link duration, average cluster lifetime, and a clusters’ existence probability between two fixed moments of time. In support of the theoretically obtained probabilistic distributions, the results of numerical simulations are provided.
\end{abstract}

\begin{IEEEkeywords}
Vehicular network, clustering, network evolution, link duration.
\end{IEEEkeywords}

\IEEEpeerreviewmaketitle

%\hfill August 4, 2015

\section{Introduction}\label{Sec:beg}
Vehicular Adhoc Networks (VANETs) have become one of the frontier topics  of research \cite{ML}--\cite{CarNet2} over the last decade due to the anticipated mass deployment of self-driving vehicles. A VANET consists of a set of fast moving vehicles equipped with sensing,   communication and infotainment systems. This turns  the neighboring connected vehicles into a moving wireless network, that enables vehicles to connect with each other and share safety and entertainment related content. Self-driving vehicles are expected to significantly reduce the number of road accidents and traffic congestion, improve road safety, and further enable intelligent transportation.

Vehicle networking is a relatively new field that allows for the application of emerging technologies such as machine learning, which is widely used in many areas from image processing to financial data analysis. Based on extensive statistics, a neural network is capable of predicting and classifying data, and therefore, it is a powerful tool for solving a variety of different problems.  Machine learning techniques have applications in optimization of information transmission between vehicles \cite{ML}, \cite{ML2}, improving transportation safety \cite{ML3}, and reducing transmission delays \cite{ML4}. The applications of machine learning in the aforementioned areas, as well as for network congestion control, wireless resource management, and load balancing are discussed in a comprehensive survey \cite{ML5}. It includes not only classical deep learning algorithms used for the system’s state prediction from available data, but also modern developments in the field of reinforcement learning, that makes it possible to find strategies leading to long-term rewards in the cases, where a problem can be reformulated as a game. Reinforcement learning algorithms demonstrate their superiority in comparison to the previous approaches in resource management and resource allocation tasks, and allow for a reduction in computational complexity.

In recent years, significant progress has been made in the area of MAC protocols for vehicular networks \cite{M1}--\cite{M6}. Due to the dynamic topology of the network, the development of robust and efficient transmission of information in Vehicle-to-Vehicle (V2V), and Vehicle-to-RSU (V2R) scenarios becomes highly significant. One important issue here is avoiding collisions where more than one node transmits at the same time interval, because resending the same packages causes delays, which is not acceptable for delay-sensitive applications. For overcoming the problem, near collision free MAC protocols adapting to different densities of traffic flow have been proposed.

MAC protocols are intrinsically complicated, which makes them almost impossible to analyze analytically from a statistical point of view. For practical applications, it is important to estimate the average cluster size of the network and probability of multihop connectivity between two vehicles. To address these challenges, in the case of a straight road (see articles \cite{P1}--\cite{My}), simplified models are proposed, making it possible to derive explicit expressions for the aforementioned fundamental network characteristics, and the probability of full connectivity. In \cite{Clust3}, the authors consider a more complicated case, where vehicle networks are moving along two perpendicular roads towards the intersection. They derive formulas for the outage probability and the transmission probability, assuming that intervehicle distance has Poisson distribution.

However, these studies are limited to a current moment, while it is important to investigate how connectivity changes over time. In particular, it is important to know what the average link duration is. In the articles \cite{CarNet} and \cite{CarNet2}, the explicit formulas for the link duration are obtained under different assumptions. In \cite{CarNet}, the average link duration is investigated assuming a two-way road, and that vehicles only transmit messages to other vehicles moving in the opposite direction. In \cite{CarNet2}, the average link duration is calculated in the case of one-way traffic under the assumption that speed increases linearly until it reaches the speed limit and then remains constant.

Network evolution is a stochastic process. Therefore, we can predict the state of the system for the next moment of time only with certain probability. It explains the application of the probability theory to the analysis of network evolution. The channel interruption probability is small, but positive, leading to an increase in the probability of disconnection over time. This article is aimed at describing the evolution of the network in terms of the probability of maintaining a connection between two consecutive vehicles and the vehicles forming a cluster, and investigating how rapidly the probability decreases. We derive explicit expressions for the probabilities of link duration, cluster existence over a certain amount of time, and the  probability of cluster existence between two fixed moments of time. The obtained results related to cluster evolution are new, and they provide an insight into dynamical changes in the network.

It is convenient to investigate the connectivity properties of the system not at every moment, but only on a discrete uniform time mesh. It means that the difference between two consecutive moments of time has a fixed duration $\Delta t$. Let $t$ be the current moment of time. The Markov model determines the probability of an event that at the moment of time $t+\Delta t$, the consecutive vehicles could establish a connection based on the communication state at the time moment $t$. This probability is explicitly expressed in terms of macro parameters of the system \cite{WM}, \cite{Primak}. Using the state-transition matrix of the Markov process, it is possible to express the desired connection probabilities in terms of the matrix coefficients, and thereby, allows the calculation of the required connectivity characteristics.

Also, we consider such a stability characteristic of the connection between two consecutive cars as $\omega$-stable connection. This type of connection guarantees that the time between consecutive connections does not exceed $\omega\Delta t$ ($\Delta t$ is a timestep). In other words, this weakened condition means that at some moments of time the vehicles may fail to establish a connection, but they connect at least once at each time interval that has length  $\omega\Delta t$. This type of communication is closer to the actual operating conditions, where connection may disappear  for short periods of time, but it is important to ensure that it is regularly reestablished. We derive recurrent equations for calculating the probabilities and verify the results using simulations.

The article is organized as follows. Section \ref{Sec:NetModel} describes the parameters and the structure of a considered network model. Section \ref{Sec:Connectivity} investigates the statistics of node-to-node connectivity between cars assuming a fading channel between the nodes.The detailed mathematical derivations are summarized in Section \ref{Sec:Derivations}, and a brief summary is given in \ref{Sec:SummaryResults}. The derived expressions are verified through a number of simulations in Sections \ref{Sec:Sim}, \ref{Sec:Sim2}.

\section{Network model}\label{Sec:NetModel}
We consider the following probabilistic model that describes the evolution of the network over time.
It is assumed that there are two states of connection \textit{Good} and \textit{Bad}.
In the \textit{Good} state, neighboring cars can establish a connection, while the \textit{Bad} state corresponds to the case where neighboring cars cannot connect with each other.
We suppose that initially the system is in the equilibrium state; the initial probabilities of connection are calculated in the section $\textrm{V}$.
We consider the system evolution process with timestep $\Delta t$ (introduced at the end of the previous section).
Let $p$ be the probability that connected neighbouring cars cannot establish a connection at the next moment of time, in other words, $p$ is the probability that at the next moment of time the system moves from the \textit{Good} state to the \textit{Bad} state.
By analogy, let $q$ be the probability that at the next moment the system moves from the \textit{Bad} state to the \textit{Good} state. Therefore, connectivity between two consecutive cars can be described by two-state Markov Model depicted below (letters \textbf{G} and \textbf{B} denote the \textit{Good} and the \textit{Bad} states, respectively). The explicit values of the parameters $p$ and $q$ are given in the next section.

\begin{center}
	\includegraphics[scale=0.15]{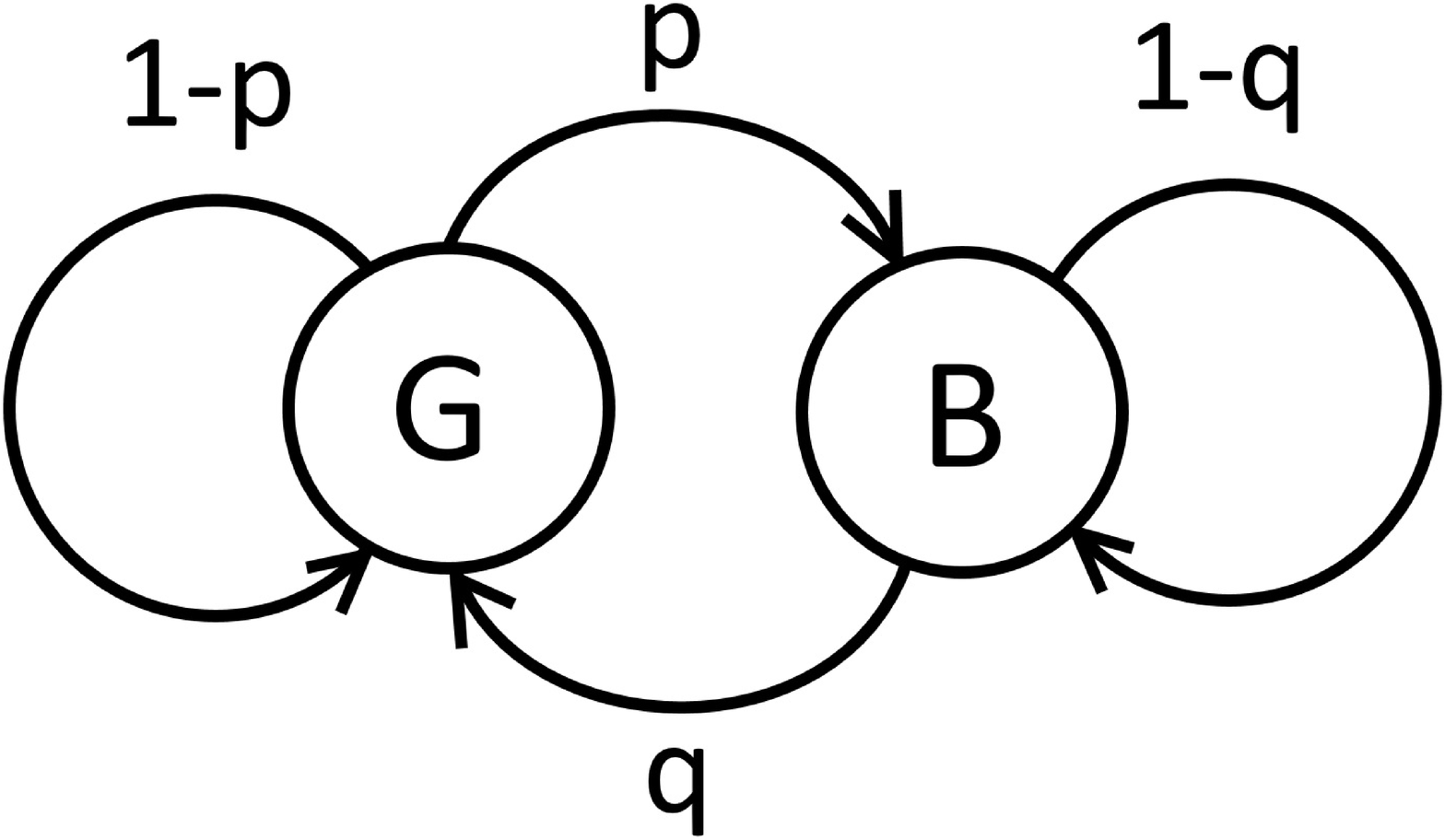}\\
	{\footnotesize Fig 1. Markov diagram of the connectivity process\\ with the \textit{Good} and the \textit{Bad} states}
\end{center}

\section{Probabilistic connectivity model}\label{Sec:Connectivity}

Everywhere in the article it is assumed that all $n$ vehicles move along one-way road with the same constant speed $v$.
Each car can establish a connection only with the closest front and back neighbours.
We assume a Rayleigh fading channel between every pair of cars. Consequently, the amplitude of the received signal is exponentially distributed with pdf
\begin{equation}
p(x)=\frac{1}{\lambda}e^{-x/\lambda},
\end{equation}
where $\lambda$ represents the average SNR (Signal-to-noise ratio) over the fading channel. We use the Wang-Moayery model \cite{Primak} to determine values of the parameters $p$ and $q$. According to this model, we determine state of the system by comparison signal amplitude $A$ and threshold $\overline{A}$. If $A\geq \overline{A}$ then we assume, that the system is in the \textit{Good} state, otherwise it is in the \textit{Bad} state. We denote by $p_B$ the probability of the \textit{Bad} case, therefore,
\begin{equation}
p_B=\int_0^{\overline{A}} p(x)dx = \int_0^{\overline{A}}\frac{1}{\lambda}e^{-x/\lambda}dx=1-e^{-\overline{A}/\lambda}.
\end{equation}

The probability of the \textit{Good} state is given by the formula
\begin{equation}
p_G=1-p_B=e^{-\overline{A}/\lambda}.
\end{equation}
Assuming the Clark model, the following formula for the level crossing rate is obtained in \cite{Primak}:
\begin{equation}
LCR(x)=\sqrt{\frac{2\pi x}{\lambda}}f_De^{-x/\lambda},
\end{equation}
where $f_D$ is Doppler shift. Also, the following two explicit formulas for parameters $p$ and $q$ (they are introduced in the previous section) are derived
\begin{equation}\label{ttt1}
p=\frac{LCR(\overline{A})}{Rp_G}=\frac{\sqrt{\frac{2\pi \overline{A}}{\lambda}}f_D}{R},
\end{equation}
\begin{equation}\label{ttt2}
q=\frac{LCR(\overline{A})}{Rp_B}=\frac{\sqrt{\frac{2\pi \overline{A}}{\lambda}}f_De^{-\overline{A}/\lambda}}{R(1-e^{-\overline{A}/\lambda})},
\end{equation}
where $R$ is the symbol rate.
Taking into account (\ref{ttt1}), (\ref{ttt2}), and the following formula for the maximum Doppler shift \footnote{Maximum Doppler shift is consistent with geometry of the problem, since vehicles follow the signal propagation path.}:
\begin{equation}\label{Doppler}
f_D=\frac{vf_c}{c},
\end{equation}
where $f_c$ and $c$ are transmitted frequency and velocity of light respectively, we obtain
\begin{equation}\label{ttt3}
p=\frac{LCR(\overline{A})}{Rp_G}=\frac{\sqrt{\frac{2\pi \overline{A}}{\lambda}}vf_c}{Rc},
\end{equation}
\begin{equation}\label{ttt4}
q=\frac{LCR(\overline{A})}{Rp_B}=\frac{\sqrt{\frac{2\pi \overline{A}}{\lambda}}vf_ce^{-\overline{A}/\lambda}}{R c(1-e^{-\overline{A}/\lambda})}.
\end{equation}
The timestep $\Delta t$ (see description of timestep in the section~\ref{Sec:beg}) satisfies Nyquist-–Shannon criterion
\begin{equation}
\Delta t\leq\frac{1}{2f_D}.
\end{equation}

\section{Main results}\label{Sec:SummaryResults}

Let $p_G(m)$ be the probability that at the moment $m\Delta t$ two fixed consecutive cars can establish a connection.
We assume that the system is initially in the equilibrium state (for more details see section $\textrm{V}$).
In the above mentioned section, the probability $p_G(m)$ of a successful connection between two consecutive cars at the moment of time $m\Delta t$ satisfies the following formula:
\begin{equation}\label{ddd11}
p_G(m)=\frac{q}{p+q},
\end{equation}
where parameters $p$ and $q$ are calculated by the formulas (\ref{ttt3}) and (\ref{ttt4}). Thus, probability $p_G(m)$ does not depend on $m$, and, therefore, we denote it by $p_G$.
%All examples below are considered under the assumption that
%\begin{equation}\label{ass}
%p=0.3,q=0.1,\Delta t=1,n=10.
%\end{equation}

%\textbf{Example 1. } The probability that cars 2 and 3 (or any other two consecutive cars) can establish a connection at the moment 5 equals to
%$$
%p_G(5)=\frac{q(1-(1-p-q )^5)}{p+q}=0.2306.
%$$

Let symbol $P_{twocars}(m)$ stand for
the probability that connection between two cars once established has duration $m\Delta t$.
The probability $P_{twocars}(m)$ is given by the formula
\begin{equation}\label{eee}
P_{twocars}(m)=(1-p)^{m-1}p.
\end{equation}

%\textbf{Example 2. } Let us find the probability that cars 2 and 3 (or any other two consecutive cars) can establish connection for exactly 3 seconds assuming (\ref{ass}). It means that the connection exists for three second and cannot be established at the forth second. This probability is given by the formula
%$$
%P_{twocars}(3)=(1-p)^{2}p=0.147.
%$$

The average link duration $\overline{T}_{twocars}$ and its variance $\sigma_{twocars}^2$ of the distribution (\ref{eee}) are determined by the formulas
\begin{equation}\label{xxxxxx}
\overline{T}_{twocars}=\frac{\Delta t}{p},
\end{equation}

\begin{equation}\label{xxxxxx}
\sigma_{twocars}^2=\frac{(1-p)\Delta t^2}{p^2}.
\end{equation}

%\textbf{Example 3. } The average link duration $\overline{T}_{twocars}$ between any two consecutive cars is given by the formula
%$$
%\overline{T}_{twocars}=\frac{1}{p}=3.3333.
%$$

\textbf{Definition. } A \textit{cluster} is a such group of vehicles that any two cars in the group can communicate with each other, probably through other vehicles of the cluster.

In our model, each vehicle connects only to the closest forward and backward cars.
Therefore, in the framework of our model, clusters are formed by several consecutive cars.
On Fig. 2 below, five cars are depicted that form two clusters. The first cluster is formed by cars 1, 2, 3 and the second cluster is composed of cars 4 and 5.
Significantly, cars 1 and 3 cannot communicate directly, and the only communication way for them is through car 2,
while vehicles 3 and 4 are disconnected. We prove that the connectivity probability is given by the formula $p/(p+q)$. Therefore, connectivity state between two consecutive vehicles depends on the parameters listed in the section \ref{Sec:Connectivity}. It is worth mentioning that this probability is close to $1$ (stable connection) if $p\approx 1$ and $q \approx 0$.

We derive the probability $P_{clust}(m)$ that the cluster formed by the cars $k,k+1,\ldots,k+s$ once being formed exists exactly time $m\Delta t$. Let us introduce parameter $\gamma$ by the formula
\begin{equation}\label{eeeii11}
\gamma=\begin{cases}
2, & \mbox{if } 1<k\mbox{ and }k+s<n, \\
1, & \mbox{if } k=1\mbox{ or }k+s=n,\mbox{ but not both, }\\
0, & \mbox{if } k=1\mbox{ and }k+s=n.
\end{cases}
\end{equation}
The probability $P_{clust}(m)$ satisfies the formula
\begin{equation}\label{eeeii}
P_{clust}(m)= (1-p)^{s(m-1)}(1-q)^{\gamma(m-1)} (1-(1-p)^s(1-q)^\gamma).
\end{equation}

\begin{center}
	\includegraphics[scale=0.2]{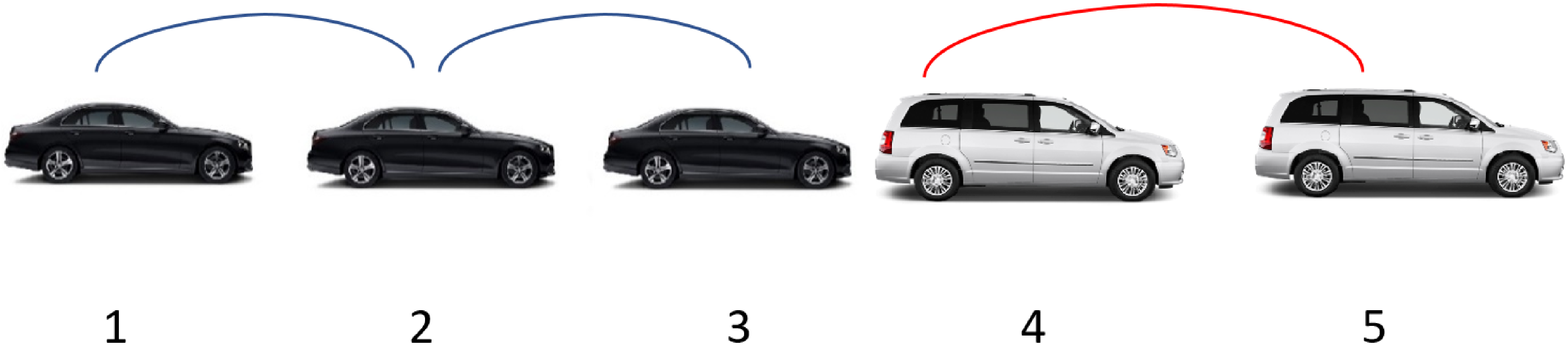}\\
	{\footnotesize Fig 2. Network of 5 cars that form two clusters.}
\end{center}
%\textbf{Example 4. } Under assumptions (\ref{ass}) let us calculate the probability that the cluster formed by the cars 2, 3 and 4 exists 3 seconds. In this case $k=2$, $s=2$ and $\gamma=2$. The probability $P_{clust}(3)$ can be calculated by the formula
%$$
%P_{clust}(3)= (1-p)^{4}(1-q)^{4} (1-(1-p)^2(1-q)^2)=0.095.
%$$

The average existence cluster lifetime $\overline{T}_{clust}$ and variance $\sigma_{clust}^2$ of the distribution (\ref{eeeii}) can be obtained by the formulas
\begin{equation}\label{rtsssr}
\overline{T}_{clust}=\frac{\Delta t}{1-(1-p)^s(1-q)^\gamma}.
\end{equation}
\begin{equation}\label{rddtr1}
\sigma_{clust}^2 =
\frac{(1-q)^\gamma(1-p)^s\Delta t^2}{(1-(1-q)^\gamma(1-p)^s)^2},
\end{equation}
where $\gamma$ is determined in (\ref{eeeii11}).

%\textbf{Example 5. } Assuming the same values of the parameters as in the example 4, we derive that average time of existence of the cluster formed by the cars 2, 3 and 4 equals to
%$$
%\overline{T}_{clust}=\frac{1}{1-(1-p)^2(1-q)^2}= 1.6581.
%$$

We study not only the duration of a cluster lifetime, but also the probability that cluster does exist between two given moments of time. More precisely, we derive the following formula for the probability $P_{clust}(m,l)$ that the cluster that consists of cars $k,k+1,\ldots,k+s$ is formed at the moment of time $m\Delta t$, exists until $l\Delta t$ ($l \geq m$), and does not exist at the moment $(l+1)\Delta t$ as follows:
\begin{multline}\label{444}
P_{clust}(m,l)=\Big\{p_G^s(1-p_G)^\gamma \Big. \\-(1-p)^s p_G^s(1-p_G)^\gamma(1-q)^\gamma\Big.\Big\}\\\times(1-p)^{s(l-m)}(1-q)^{\gamma(l-m)}\Big\{1-(1-p)^s(1-q)^\gamma\Big\},
\end{multline}
where $p_G$ and $\gamma$ are given by the formulas (\ref{ddd11}) and (\ref{eeeii11}).

%\textbf{Example 6. } Assuming (\ref{ass}) let us calculate the probability that the cluster consisnting the cars 2, 3 and 4
%exists exactly between moments $m=2$ and $l=4$. Substituting $s=k=\gamma=2$ into the formula (\ref{444}) we conclude that
%$$
%P_{clust}(2,4)= 1.4\times{10^{-3}}.
%$$

Finally, we analyse the property of $\omega$-stability of a connection
that is defined as the ability of two vehicles to establish connection within every time interval $\omega\Delta t$.
More formally, we assume that connection is $\omega$-stable between moments $m\Delta t$ and $l\Delta t$,
if there is at least one successful connection not later that the moment $(m+\omega)\Delta t$, time difference between every two consecutive connections does not exceed $\omega\Delta t$ and the last connection is established at the moment $(l-\omega)\Delta t$ or later.
At the end of the section $\textrm{V}$, we derive the algorithm for  finding the  probability that the connection between two cars is $\omega$-stable on a pre-selected interval.
The algorithm has linear complexity and recurrently calculates the probability under the assumption of known parameters $p$ and $q$.

\section{Mathematical derivations}\label{Sec:Derivations}

\subsection{Probability of two cars being connected at the moment $m\Delta t$}

Let $p_G(m)$ be the probability that there is a connection between predetermined consecutive cars at the moment of time $m\Delta t$ and by $p_B(m)$ the probability that there is no connection at $m\Delta t$.
Let us find the limiting distributions $p_G(\infty)=\lim_{m\to\infty}p_G(m)$ and $p_B(\infty)=\lim_{m\to\infty}p_B(m)$.
According to the Markov model (see Fig. 1), we have
\begin{equation}\label{ddd12}
p_G(m) = p_G(m-1)(1-p) + p_B(m-1)q,
\end{equation}
\begin{equation}\label{ddd13}
p_G(m) + p_B(m) = 1.
\end{equation}
When $m \to\infty$, the equations (\ref{ddd12}) and (\ref{ddd13}) take the following form
\begin{equation}\label{dd12}
p_G(\infty) = p_G(\infty)(1-p)+p_B(\infty)q,
\end{equation}
\begin{equation}\label{dd13}
p_G(\infty) + p_B(\infty)=1.
\end{equation}
Solving system of linear equations (\ref{dd12}) and (\ref{dd13}), we obtain
\begin{equation}\label{limpG}
p_G(\infty)=\frac{q}{p+q},
\end{equation}
\begin{equation}\label{limpB}
p_B(\infty)=\frac{p}{p+q}.
\end{equation}

We assume that at the moment when we observe the system, it has reached its limiting distribution, therefore,  the probabilities at every moment $m\Delta t$ are equal to the limiting probabilities:
%\begin{equation}\label{init}
%p_G(0)=\frac{q}{p+q},
%\end{equation}
%\begin{equation}\label{init2}
%p_B(0)=\frac{p}{p+q}.
%\end{equation}
%
%%If we substitute $m=1$ in the formulas (\ref{ddd12}) and (\ref{ddd13}) and solve the system of linear equations we get that $p_G(1)=p_G(0)$ and $p_B(1)=p_B(0)$. Continuing this process further we obtain that
%
%Taking into account that the limiting distribution has the
\begin{equation}\label{prob}
p_G(m)=\frac{q}{p+q},
\end{equation}
\begin{equation}\label{prob2}
p_B(m)=\frac{p}{p+q}.
\end{equation}
Thus, we denote them $p_G$ and $p_B$.

%Further in the paper, we assume that the instantaneous probabilities $p_G(m)$ and $p_B(m)$ at each moment of time $m\Delta t$ does not depend on $m$ and equal to the limiting probabilities $P_G(\infty)$, $i\in\{G, B\}$, \textit{i.e.}}

%\begin{equation}\label{init}
%p_G(0)=\frac{q}{p+q},
%\end{equation}
%\begin{equation}\label{init2}
%p_B(0)=\frac{p}{p+q}.
%\end{equation}
%
%%If we substitute $m=1$ in the formulas (\ref{ddd12}) and (\ref{ddd13}) and solve the system of linear equations we get that $p_G(1)=p_G(0)$ and $p_B(1)=p_B(0)$. Continuing this process further we obtain that
%
%Taking into account that the limiting distribution has the

%
%
%The probabilities $p_G(k)$ and $p_B(m)$ can be calculated by using the transition matrix by the following well-known formula:
%\begin{equation}\label{11}
%\begin{bmatrix}
%           p_G(r) &
%           p_B(r)
%\end{bmatrix}=\begin{bmatrix}
%           p_G(0) & p_B(0)
%\end{bmatrix}\times
%\begin{bmatrix}
%           1-p & p \\
%           q & 1-q
%\end{bmatrix}^r.
%\end{equation}
%By using the following identity for the power of the transition matrix
%$$
%\begin{bmatrix}
%           1-p & p \\
%           q & 1-q
%\end{bmatrix}^r=
%\begin{bmatrix}
%\frac{p(1-p-q)^r+q}{p+q} & \frac{p(1-(1-p-q)^r)}{p+q}\\
%\frac{q(1-(1-p-q )^r)}{p+q} & \frac{(1-p-q )^r q+p}{p+q}
%\end{bmatrix}
%$$
%and formulas (\ref{init}), (\ref{init2}) and (\ref{11}) we derive that
%\begin{equation}\label{ddd}
%p_G(r)=\frac{q(1-(1-p-q )^r)}{p+q},
%\end{equation}
%\begin{equation}\label{ddd1}
%p_B(r)=\frac{(1-p-q )^r q+p}{p+q}.
%\end{equation}

\subsection{Distribution of link duration}

Let us consider two consecutive cars.
We find the probability $P_{twocars}(m)$ of the event that the connection lifetime between these cars is exactly $m\Delta t$.
In other words, if the connection between cars is firstly established at the moment of time $\Delta t$, we calculate the probability that the cars keep the connection up to the time instant $m\Delta t$, and at the time $(m+1)\Delta t$ the connection cannot be established.
This probability is given by the formula
\begin{equation}\label{rrrrr34}
P_{twocars}(m)=(1-p)^{m-1}p,
\end{equation}
since at the time $\Delta t$ the system is in the \textit{Good} state, and the probability that it remains in the \textit{Good} state at the moments $2\Delta t$, $3\Delta t,\ldots m\Delta t$ is $(1-p)^{m-1}$ (see Markov diagram on {Fig. 1}), and the probability that the system changes the state from {\it Good} to {\it Bad} at the time instant $(m+1)\Delta t$ is $p$.

Using relation~\eqref{rrrrr34}, we can calculate the average link duration $\overline{T}_{twocars}$ and variance $\sigma_{twocars}^2$ of the distribution by the formulas
\begin{equation}\label{yyy}
\overline{T}_{twocars}=\sum_{l=1}^\infty l\Delta t P_{twocars}(l)=\sum_{l=1}^\infty l\Delta t (1-p)^{l-1}p=\frac{\Delta t}{p},
\end{equation}
\begin{multline}\label{222}
\sigma_{twocars}^2=\sum_{l=1}^\infty (l\Delta t)^2 P_{twocars}(l)-\overline{T}_{twocars}^2\\=\sum_{l=1}^\infty l^2 \Delta t^2 (1-p)^{l-1}p-\frac{\Delta t^2}{p^2}\\=\Big(\frac{2-p}{p^2}-\frac{1}{p^2}\Big)\Delta t^2=\frac{(1-p)\Delta t^2}{p^2}.
\end{multline}

%$P_{twocars}(r,t)$ that connection between them is established at $r\Delta t$ and exist until $t\Delta t$. This probability is given by the formula
%$$
%P_{twocars}(r,t)=(1-p_G(r-1))q(1-p)^{t-r}p,
%$$
%because probability that cars do not connect at the time $(r-1)\Delta t$ equals to $(1-p_G(r-1))$, where $p_G(r)$ is given by the formula (\ref{ddd}), the probability that the system move from state \textit{Bad} at the moment $(r-1)\Delta t$ to state \textit{Good} at the moment $r\Delta t$ equals to $q$ (see Markov diagram), the probability that the system stay in state \textit{Good} from the moment $r\Delta t$ to $t\Delta t$ is $(1-p)^{r-s}$ and finally the probability that the system move from the \textit{Good} state at the moment $t\Delta t$ to \textit{Bad} state at the moment $(t+1)\Delta t$ is p.

\subsection{Distribution of a cluster lifetime}

In this section we find the probability $P_{clust}(m)$ that the existence duration of a cluster composed by cars $k$, $k+1,\ldots,k+s$ equals exactly $m\Delta t$. We could assume that the cluster exists since the moment $\Delta t$ until the moment $m\Delta t$ and does not exist at the moment $(m+1)\Delta t$.
The probability $P_{clust}(m)$ satisfies the formula
\begin{equation}\label{eeee}
P_{clust}(m)= (1-p)^{s(m-1)}(1-q)^{\gamma(m-1)} (1-(1-p)^s(1-q)^\gamma),
\end{equation}
where $\gamma$ is defined by (\ref{eeeii11}).

Let us derive the equation (\ref{eeee}).
We consider only the case where $k > 1$ and $k+s < n$, because in other cases derivations are similar. The probability $P_{clust}(m)$ can be calculated by the formula
\begin{equation}\label{eeee1211}
P_{clust}(m)= P_1 P_2,
\end{equation}
where $P_1$ is a probability that the cluster exists at the moments $2\Delta t, \ldots m\Delta t$ under the condition that it exists at the moment $\Delta t$, and $P_2$ is a probability that the cluster does not exist at the moment $(m+1)\Delta t$ under the assumption that it exists at the moment $m\Delta t$. First, we  calculate the probability $P_1$.
The fact that the cluster $k,k+1,\ldots,k+s$ exists at the moment $\Delta t$ simply means that at this moment there is a connection between every pair of cars $(k,k+1)$, ${(k+1,k+2)}$, $\ldots$, ${(k+s-1,k+s)}$ and there is no connection between pairs of cars $(k,k-1)$ and $(k+s,k+s+1)$.
The probability that all pairs of cars  $(k,k+1)$, ${(k+1,k+2)}$, $\ldots$, ${(k+s-1,k+s)}$ preserve a connection at the moment $2\Delta t$ is $(1-p)^s$ (see Markov diagram on Fig. 1).
The probability that the pairs of cars $(k,k-1)$ and $(k+s,k+s+1)$ continue to be disconnected at the moment $2\Delta t$ is $(1-q)^2$. Thus, the probability that the cluster exists at the moment $2\Delta t$ is $(1-p)^s(1-q)^2$. Continuing these derivations until the moment $m\Delta t$, we find the probability $P_1$ given as follows:
\begin{equation}\label{eeee121111}
P_1 =(1-p)^{s(m-1)}(1-q)^{2(m-1)},
\end{equation}
since there are $m-1$ transitions between steps $\Delta t$ and $m\Delta t$ and at each step we need to multiply the answer by the transitional probability $(1-p)^s(1-q)^2$.
Finally, we derive the probability $P_2$.
The cluster should cease to exist at the moment $(m+1)\Delta t$.
The probability that the cluster continues to exist at the next moment is $(1-p)^s(1-q)^2$, therefore, the probability that it does not exist is
\begin{equation}\label{eeee22121111}
P_2=1-(1-p)^s(1-q)^2.
\end{equation}
Formulas (\ref{eeee1211})--(\ref{eeee22121111}) prove the equation (\ref{eeee}).

The average time of cluster existence $\overline{T}_{clust}$ and its variance $\sigma_{clust}^2$ are determined by the following formulas:
\begin{multline}\label{rtr}
\overline{T}_{clust}=\sum_{l=1}^\infty l\Delta t P_{clust}(l)\\=\sum_{l=1}^\infty l\Delta t(1-p)^{s(l-1)}(1-q)^{\gamma(l-1)} (1-(1-p)^s(1-q)^\gamma)\\=
\frac{\Delta t}{1-(1-p)^s(1-q)^\gamma}.
\end{multline}
\begin{multline}\label{rtr1}
\sigma_{clust}^2=\sum_{l=1}^\infty (l\Delta t)^2P_{clust}(l) - \overline{T}_{clust}^2\\=\frac{(1+(1-p)^s(1-q)^\gamma)\Delta t^2}{(1-(1-p)^s(1-q)^\gamma)^2}-\Big(\frac{\Delta t}{1-(1-p)^s(1-q)^\gamma}\Big)^2\\=
\frac{(1-q)^\gamma(1-p)^s\Delta t^2}{(1-(1-q)^\gamma(1-p)^s)^2}.
\end{multline}

\subsection{Probability of cluster existence between fixed moments of time}

%Let us start with computing the probability $P_{clustbeg}(a)$ that at the moment $a\Delta t$ there is cluster consisting of cars $k,k+1$, $\ldots, k+s$.
In the section, we derive the probability of a cluster existence between two particular moments of time.
We consider only the case where $k>1$ and $k+s<n$, because in other cases derivations are similar.
However, before solving the problem, we find the probability $P_{clustbeg}(m)$ that at the time $m\Delta t$ the cluster that consists of cars $k,k+1,\ldots, k+s$, is formed, and it does not exist at the moment $(m-1)\Delta t$.
%Taking this into account, we derive that the probability $P_{clustbeg}(m)$ is given by the formula \Cr{[???? You missed something here? About what formula do you write here?]}
Let us denote the probability of the event that at the moment $m\Delta t$ there is a cluster consisting of cars $k,k+1$, $\ldots, k+s$ by $P_1$ and the probability that the cluster $k,k+1$, $\ldots, k+s$ exists at the moments $(m-1)\Delta t$ and $m\Delta t$ by $P_2$. Therefore, the probability $P_{clustbeg}(m)$ is given as follows:
\begin{equation}\label{er1}
P_{clustbeg}(m)=P_1-P_2.
\end{equation}
Next, we  calculate each of the probabilities $P_1$ and $P_2$ separately.
Firstly, we establish that
\begin{equation}\label{er}
P_1=p_G^s(1-p_G)^2.
\end{equation}
Indeed, if the cluster $k,k+1$, $\ldots, k+s$ exists at the moment $m\Delta t$ then
the pairs of cars $(k,k+1)$, $(k+1,k+2),\ldots,(k+s-1,k+s)$ are connected (each with the probability $p_G$). Thus, the probability of this event is $p_G^s$, since there are $s$ such pairs.
Additionally, there is no connection between the pairs of cars $(k,k-1)$ and $(k+s$, $k+s+1)$ (each of disconnections occurs with the probability $1-p_G$).
Multiplying the aforementioned probabilities, we get the relation (\ref{er}).
Regarding the probability $P_2$, we prove the formula
\begin{equation}\label{er2}
P_2=(1-p)^s p_G^s(1-p_G)^2(1-q)^2.
\end{equation}
The probability $P_2$ can be represented as follows:
\begin{equation}\label{er333}
P_2=P_3 P_4,
\end{equation}
where $P_3$ is a probability that the cluster $k,k+1, \ldots, k+s$ exists at the moment $(m-1)\Delta t$ and $P_4$ is a probability that it continues to exist at the moment of time $m\Delta t$ under the assumption that it exists at the moment $(m-1)\Delta t$.
Repeating derivations for the probability $P_1$ in the case of the  time instant $(m-1)\Delta t$, we have
\begin{equation}\label{er22222}
P_3=P_1=p_G^s(1-p_G)^2.
\end{equation}
Next, we prove the relation
\begin{equation}\label{er2333}
P_4=(1-p)^s(1-q)^2.
\end{equation}
Suppose, the cluster exists at the moment of time $(m-1)\Delta t$.
The probability that a connection between any pair of the consecutive cars of the cluster is preserved at the moment $m\Delta t$ equals $1-p$.
Consequently, the probability that the connection between all $s$ pairs of cars is preserved equals $(1-p)^s$.
The pairs of cars $(k-1,k)$ and $(k+s,k+s+1)$ cannot establish a connection at the moment $(m-1)\Delta t$, the probability that these pairs remain disconnected is $(1-q)^2$. Multiplying two probabilities $(1-p)^s$ and $(1-q)^2$, we derive the formula (\ref{er2333}).

From (\ref{er})--(\ref{er2333}), we conclude that
\begin{multline}\label{rr}
P_{clustbeg}(m)=p_G^s(1-p_G)^2\\-(1-p)^s p_G^s(1-p_G)^2(1-q)^2.
\end{multline}

Using the obtained probabilities, we can find the probability $P_{clust}(m,l)$ that the cluster $k,k+1,\ldots k+s$ exists between moments of time $m\Delta t$ and $l\Delta t$ $(l \geq m)$. In other words, it is the probability that the cluster exists from the time instant $m\Delta t$ to $l\Delta t$ $(l \geq m)$, and does not exist at the moments $(m-1)\Delta t$ and $(l+1)\Delta t$.
The probability $P_{clust}(m,l)$ satisfies the formula
\begin{multline} \label{324234}
P_{clust}(m,l)=(p_G^s(1-p_G)^2\\-(1-p)^s p_G^s(1-p_G)^2(1-q)^2)\\ \times(1-p)^{s(l-m)}(1-q)^{2(l-m)}(1-(1-p)^s(1-q)^2).
\end{multline}
The probability $P_{clust}(m,l)$ can be decomposed in the product of three terms as follows:
\begin{equation} \label{3242341}
P_{clust}(m,l) = P_{clustbeg}(m)P_5P_6,
\end{equation}
where $P_5$ is a probability that the cluster exists until the moment $l\Delta t$, and $P_6$ is a probability that the cluster ceases to exist at the moment $(l+1)\Delta t$ assuming that it exists at the moment $l\Delta t$. First, we consider the probability $P_5$ and prove the formula
\begin{equation} \label{32423334}
P_5 = (1-p)^{s(l-m)}(1-q)^{2(l-m)}.
\end{equation}
By analogy with derivations of (\ref{eeee121111}), we deduce that the probability, that the connection established between the pairs of cars $(k,k+1)$, $(k+1,k+2),\ldots,(k+s-1,k+s)$ at the moment of time $m\Delta t$ is preserved at the time instances $(m+1)\Delta t,\ldots ,l\Delta t$, equals $(1-p)^{s(l-m)}$.
Following the same logic of \eqref{eeee121111}, we derive that the probability, that the pairs of cars $(k-1,k)$ and $(k+s,k+s+1)$ do not communicate at the moments of time $(m+1)\Delta t,\ldots , l\Delta t$ under the condition that the connection is not established at the moment $(m-1)\Delta t$ equals $(1-q)^{2(l-m)}$.
The probability $P_5$ can be calculated as a product of these probabilities and, therefore, (\ref{32423334}) is proven.
By repeating steps of the proof of the equation (\ref{eeee22121111}), we conclude that
\begin{equation} \label{32433323334}
P_6=1-(1-p)^s(1-q)^2.
\end{equation}
From the formulas (\ref{rr}), (\ref{3242341})--(\ref{32433323334}), we derive (\ref{324234}).

\subsection{$\omega$-stable connection} \label{Sec:StCon}
\textbf{Definition}\quad
A connection between moments of time $m$ and $l$ ($m \leq l$) is $\omega$-stable
if the time difference between every two consecutive connections does not exceed time $\omega\Delta t$.
Additionally we assume that there exists at least one successful connection established not later than $(m + \omega)\Delta t$, and the last connection is established at the moment of time $(l-\omega)\Delta t$ or later.

It does not make sense to consider $\omega$-stable connection if $l<m+\omega$, because in this case every connection is $\omega$-stable.
Therefore, we assume that $l\geq m+\omega$.
Also, we suppose that $\omega$ is an integer number and $\omega\geq 2$.

In the section, we find the probability $P_\omega(m,l)$ that the connection between two consecutive cars is $\omega$-stable between the moments $m\Delta t$ and $l\Delta t$.
Let us introduce the function $h(a,b)$ equaling the probability that at the moment of time $a\Delta t$, the last connection is established at the time instant $b\Delta t$, and
the function $g(a)$ equaling the probability that there is no connection at the time interval $[0, a\Delta t]$.
To derive the function $g(a)$ explicitly, we should multiply the probability $(1-p_G)$, that at the moment $m\Delta t$ there is no connection, by the probability $(1-q)^{a-m}$ that at the moments $(m+1)\Delta,\ldots a\Delta$ there is no connection as well.
Therefore, the function $g(a)$ has the form
\begin{equation} \label{gg}
g(a)=\begin{cases}
(1-p_G)(1-q)^{a-m}, & \mbox{if } a<m+\omega \\
0, & \mbox{if } a\geq m+\omega.
\end{cases}
\end{equation}
The following recurrences take place:
%\begin{equation}\label{rrr0}
%h(a,b)=0, b<m,
%\end{equation}
\begin{equation}\label{rrr}
h(a,b)=0, b<a-\omega,
\end{equation}
\begin{equation}\label{rrr1}
h(a,b)=h(a-1,b)(1-q), a-\omega\leq b<a-1,
\end{equation}
\begin{equation}\label{rrr2}
h(a,a-1)=h(a-1,a-1)p,
\end{equation}
\begin{equation}\label{rrr3}
h(a,a)=h(a-1,a-1)(1-p)+\sum_{r=a-\omega}^{a-2} h(a-1,r)q+g(a-1)q.
\end{equation}
The formula (\ref{rrr}) holds, since for $b<a-\omega$, the probability $h(a,b)$ equals zero due to the fact that the time difference between the last connection at the moment of time $b\Delta t$ and the next connection exceeds $\omega\Delta t$.
The formula (\ref{rrr1}) holds, because if $b<a-1$ then at the moment $(a-1)\Delta t$ the system is in the \textit{Bad} state, and the probability that it remains in this state is $1-q$.
The formula (\ref{rrr2}) is derived by applying the same logic.
Finally, in (\ref{rrr3}), the recursive formula for the probability that the connection is established at the time instant $a\Delta t$, is obtained.
Due to the $\omega$-stability of the connection, the previous time of the connection lies between $(a-\omega)\Delta t$ and $(a-1)\Delta t$.
The summand $h(a-1,a-1)(1-p)$ occurs in the formula (\ref{rrr3}), since it is the probability that at the moment of time $(a-1)\Delta t$ the system is in the \textit{Good} state and remains at this state up to the moment of time $a\Delta t$.
The sum $\sum_{r=a-\omega}^{a-2} h(a-1,r)q$ contains probabilities $h(a-1,r)$ that the time instant $r\Delta t$ is the last moment of time when the system is in the \textit{Good} state.
It means that the system is in the \textit{Bad} state at the moment $(a-1)\Delta t$ and moves to the \textit{Good} state at the moment $a\Delta t$. It explains the multiplication of the terms $h(a-1,l)$ by $q$.

The probability $P_\omega(m,l)$ can be calculated by the formula
\begin{equation}\label{pp}
P_\omega(m,l)=\sum_{r=l-\omega}^l h(l,r),
\end{equation}
because the last moment of time before $l\Delta t$ when the system is in the \textit{Good} state, can be only $l\Delta t,(l-1)\Delta t,\ldots$, $(l-\omega)\Delta t$.

The following formulas hold:
\begin{equation}\label{in}
h(m,m)=p_G,
\end{equation}
\begin{equation}\label{in2}
h(m+1,m+1)=p_G,
\end{equation}
where $p_G$ is determined by the formula (\ref{limpG}).
The equality (\ref{in}) takes place , since the probability that connection is established at the moment $m\Delta$ is exactly $p_G$, the formula (\ref{in2}) is obtained similarly, taking into account that $\omega\geq 2$.

Under condition of $a > b$, application of the relations (\ref{rrr1}) and (\ref{rrr2}) provides the expression
\begin{multline}\label{ttttt}
h(a,b)=h(a-1,b)(1-q)=h(a-2,b)(1-q)^2=\ldots \\
=h(b+1,b)(1-q)^{a-b-1}=h(b,b)(1-q)^{a-b-1}p.
\end{multline}
From (\ref{rrr3}) and (\ref{ttttt}), we obtain
\begin{multline}\label{ttttt1}
h(a,a)=h(a-1,a-1)(1-p)\\+\sum_{r=a-\omega}^{a-2} h(a-1,r)q +g(a-1)q\\=
h(a-1,a-1)(1-p)  \\+\sum_{r=a-\omega}^{a-2} h(r,r)(1-q)^{a-r-2}pq+g(a-1)q.
\end{multline}
By substituting $a-1$ instead of $a$ into (\ref{ttttt1}), we derive
\begin{multline}\label{ttttt3333}
h(a-1,a-1)\\=h(a-2,a-2)(1-p)  \\+\sum_{r=a-\omega-1}^{a-3} h(r,r)(1-q)^{a-r-3}pq+g(a-2)q.
\end{multline}
Multiplying (\ref{ttttt3333}) by $(1-q)$, we get the following equality:
\begin{multline}\label{ttttt2}
h(a-1,a-1)(1-q)=h(a-2,a-2)(1-p)(1-q) \\+ \sum_{r=a-\omega-1}^{a-3} h(r,r)(1-q)^{a-r-2}pq+g(a-2)q(1-q).
\end{multline}
Subtraction (\ref{ttttt2}) from (\ref{ttttt1}) gives us
\begin{multline}\label{ttttt333333}
h(a,a)-h(a-1,a-1)(1-q)=h(a-1,a-1)(1-p)\\-h(a-2,a-2)(1-p)(1-q) +  h(a-2,a-2)pq \\-h(a-\omega-1,a-\omega-1) (1-q)^{\omega-1}pq\\+g(a-1)q-g(a-2)q(1-q).
\end{multline}
After simplifying (\ref{ttttt333333}), we finally derive
\begin{multline}\label{rrrrrr}
h(a,a)=h(a-1,a-1)(2-p-q)\\-h(a-2,a-2)(1-p-q) \\-h(a-\omega-1,a-\omega-1) (1-q)^{\omega-1}pq\\+g(a-1)q-g(a-2)q(1-q).
\end{multline}

Let us introduce the function $f(a)$ by the formula
\begin{equation}\label{ff}
f(a)=h(a,a).
\end{equation}
Expressions (\ref{rrrrrr}) and (\ref{ff}) provide the relation
\begin{multline}\label{ttttii}
f(a)=f(a-1)(2-p-q)-f(a-2)(1-p-q)\\ -f(a-\omega-1) (1-q)^{\omega-1}pq+g(a-1)q-g(a-2)q(1-q).
\end{multline}

Our next goal is to express the probability $P_\omega(m,l)$ in terms of the function $f(a)$.
Substitution $a=l+1$ in (\ref{rrr3}) gives us
\begin{equation}\label{ewrr}
h(l+1,l+1)=h(l,l)(1-p)+q\sum_{r=l+1-\omega}^{l-1} h(l,r)+g(l)q.
\end{equation}
Subtracting (\ref{pp}) multiplied by $q$  from (\ref{ewrr}) produces the following equality:
\begin{multline}\label{232}
h(l+1,l+1)- qP_\omega(m,l)\\
= h(l,l)(1-p)-h(l,l)q-h(l,l-\omega)q+g(l)q.
\end{multline}
Applying (\ref{ttttt}) to (\ref{232}), we get
\begin{multline}
h(l+1,l+1)- qP_\omega(m,l)=h(l,l)(1-p-q) \\-  h(l-\omega,l-\omega)(1-q)^{\omega-1}pq+g(l)q.
\end{multline}

Therefore, taking into account (\ref{ff}), we derive
\begin{multline}\label{eca}
P_\omega(m,l)=\frac{1}{q}\Big\{f(l+1)-f(l)(1-p-q)\\+  f(l-\omega)(1-q)^{\omega-1}pq-g(l)q\Big\}.
\end{multline}
From (\ref{in}) and (\ref{in2}), we obtain
\begin{equation}\label{eca2}
f(m)=p_G, \ f(m+1)=p_G.
\end{equation}

Combining previous formulas we derive the algorithm below for
calculating the probability $P_\omega(m,l)$.
In lines 2, 3, 4 and 7 two arrays $f$ and $g$ are declared and initialized to zero.
Lines 5, 6 and 9 are obtained by the formulas (\ref{eca2}) and (\ref{gg}), respectively.
Lines 12--14 are derived by (\ref{ttttii}), the return value in the line 17 is obtained by the formula (\ref{eca}).

\begin{algorithm}
%\caption*{\textbf{Algorithm. }Function for calculating the probability $P_\omega(m,l)$}
\caption*{Function for calculating the probability $P_\omega(m,l)$}
\begin{algorithmic}[1]
\Function{Probability of $\omega$-stable connection}{$m,l,\omega,p,q$}
\State$f=ARRAY[1..l+1];$
\State $g=ARRAY[1..l+1];$
\State $f=\mathbf{0};$
\State $f(m)=\frac{q}{p+q};$
\State $f(m+1)=\frac{q}{p+q};$
\State $g=\mathbf{0};$
\For{$a = m$ to $m+\omega-1$}
\State $g(a)=\frac{p}{p+q}(1-q)^{a-m};$
\EndFor
\For{$a = m+2$ to $l+1$}
\State\begin{multline*}
f(a)=f(a-1)(2-p-q)\\-f(a-2)(1-p-q)\\+g(a-1)q-g(a-2)q(1-q); \end{multline*}
\If{$a-\omega-1\geq m$}
\State$f(a)=f(a)-f(a-\omega-1) (1-q)^{\omega-1}pq;$\EndIf
\EndFor
\State \textbf{return}
\begin{multline*} \frac{1}{q}\Big\{f(l+1)-f(l)(1-p-q) \\+ f(l-\omega)(1-q)^{\omega-1}pq-g(l)q\Big\};\end{multline*}
\EndFunction
\end{algorithmic}

\end{algorithm}

\section{Simulations}\label{Sec:Sim}

In order to verify the theoretical development we conduct a number of numerical simulations with the following parameters:
%The simulations are made under the following assumptions
$$
R=10^5 \text{symbol/sec}, fc=3.9 \text{GHz}, \overline{A}/\lambda=0.1.
$$
The speed of cars $v$ takes the following values:
$$
v=30, 60, 90\text{ km/h}.
$$
The Doppler frequency shift is calculated by the formula
$$
f_D=\frac{vf_c}{c}.
$$
For the speeds $v=30, 60, 90\text{ km/h}$ Doppler frequency $f_D$ equals $108,217,325 \text{ Hz}$, respectively.
The timestep $\Delta t$ between two consecutive transmissions should satisfy the Nyquist--Shannon criteria
$$
\Delta t\leq\frac{1}{2f_D}.
$$
We assume that $\Delta t$ is given by the formula
$$
\Delta t=\frac{1}{10f_D}.
$$
For speeds $v=30, 60, 90\text{ km/h}$, values of the parameter $\Delta t$ are $9\times 10^{-4},4.6\times 10^{-4},3\times 10^{-4}$ seconds, respectively.
%Also we assume that
%$$
%\frac{\overline{A}}{\lambda}=0.1
%$$
Under these assumptions from (\ref{ttt1}) and (\ref{ttt2}), we get the following values of parameters $p$ and $q$ for speeds $v=30, 60, 90\text{ km/h}$:
$$
p=8.5\times 10^{-4},1.7\times 10^{-3}, 2.5\times 10^{-3},
$$
$$
q=8.1\times 10^{-3}, 1.6\times 10^{-2}, 2.5\times 10^{-2}.
$$
We assume that the network consists of $n=10$ vehicles.

%The probability decreasing on logarithmic scale looks linear, therefore, it means an exponential decreasing in the standard scale.

%We make simulations confirming validity of all probability distributions derived in the previous sections.

We use logarithmic scale for all graphs below.
We consider the probabilities of the event that the system remains connected throughout the time interval $[0, t]$.
By this, we mean that the connection is established at each moment of time on a discrete time grid during the interval $[0, t]$. If we increase the value of the parameter $t$, then we  increase the number of time moments when the vehicles should remain connected, so the probability of this event decreases. In other words, the considered probabilities are monotonically decreasing functions of the parameter $t$.
%\Cr{The graph \Cb{Fig. \#} illustrates the described dependencies.}
All graphs are represented as straight lines, since the probabilities exponentially decreasing with time, which in logarithmic scale, is represented as linear decreasing straight lines.
We compare the results obtained by the formulas and the results of numerical simulations.
The simulation results obtained by generating evolution of the vehicle network on the road many times, calculating number of the cases where  required characteristics of the network occur and dividing this number by the number of trials.
The values of parameters $p$ and $q$ are too small (about $10^{-3}$--$10^{-4}$) that results in negligible probabilities computed by the formulas (\ref{eee}), (\ref{eeeii}) and (\ref{444}). It makes impossible to achieve an appropriate simulation precision for the reasonable time.
Therefore, we do not depict the corresponding simulation results on the Figs 3, 4 and 5.
%showing graphs of probabilities defined by the formulas (\ref{eee}), (\ref{eeeii}) and (\ref{444}).

Fig. 3 presents graphs of the distribution (\ref{eee}) of the link duration between two consecutive cars for $v=30,60,90$ km/h.

\begin{center}%log
	\includegraphics[scale=0.54]{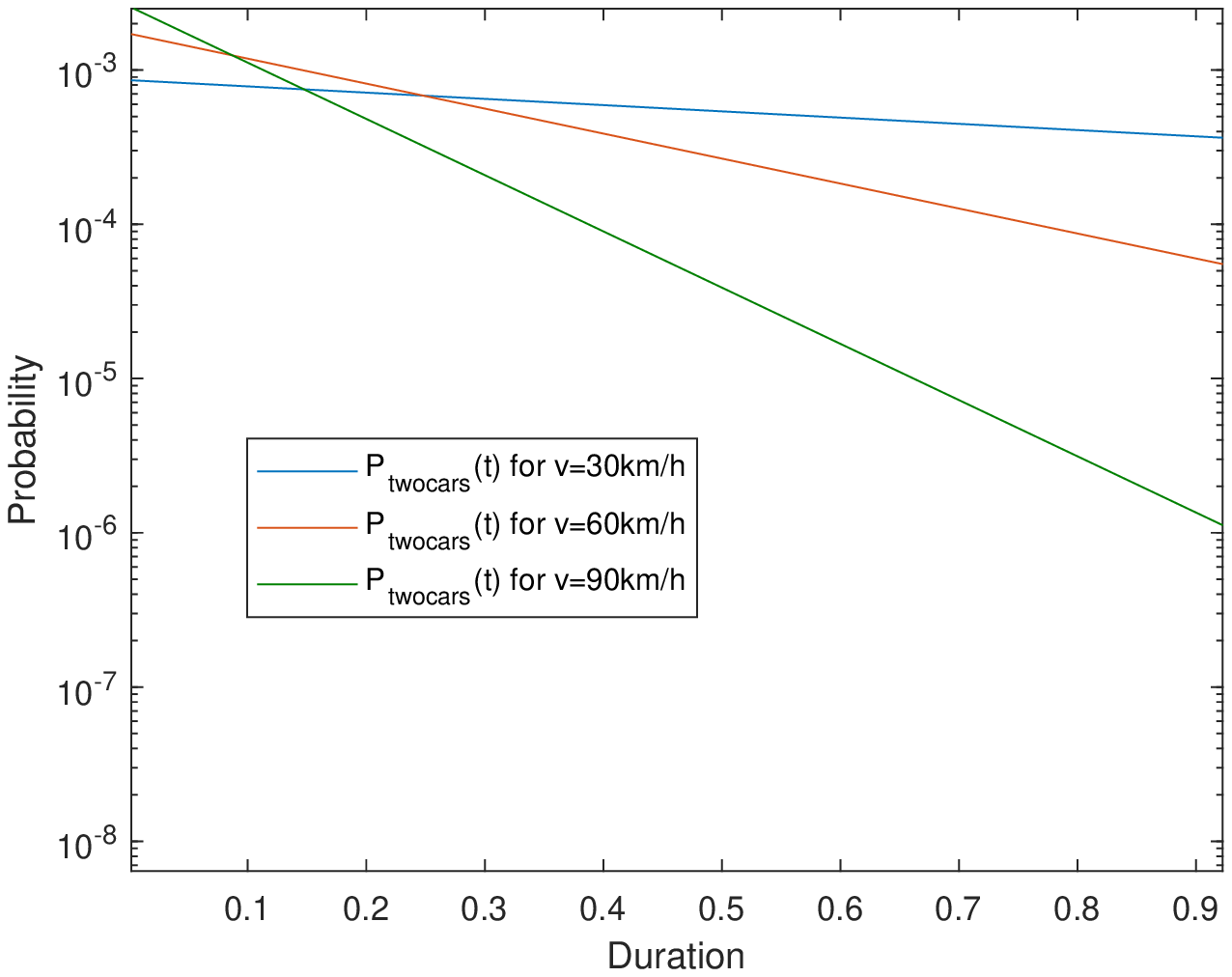}\\
	{\footnotesize Fig 3. Graphs of the distribution (\ref{eee}) of the link duration between two consecutive cars for $v=30,60,90$ km/h}
\end{center}

Fig. 4 shows the distribution (\ref{eeeii}) of the lifetime of the cluster that consists of cars 2, 3 and 4 for $v=30,60,90$ km/h.

To illustrate decreasing of the probability (\ref{444}) of the cluster existence between fixed moments of time $m\Delta t$ and $l\Delta t$, we fix the value of the parameter $m=2$ and vary the value of $l=m,m+1,\ldots$.
We assume that the cluster consists of cars with numbers 2, 3 and 4.
The graphs of the obtained functions of parameter $l$ are shown on the Fig. 5 below.

Fig.~6 demonstrates both the simulation results and the probabilities computed by the algorithm from the section \ref{Sec:StCon}.
We fix an initial moment of time $m\Delta t$ and assume that $m=2$, and vary $l=m,m+1,\ldots$.
Simulations results are obtained after 50000 generations of the car evolution process.

\begin{center} %new2_checked2
	\includegraphics[scale=0.54]{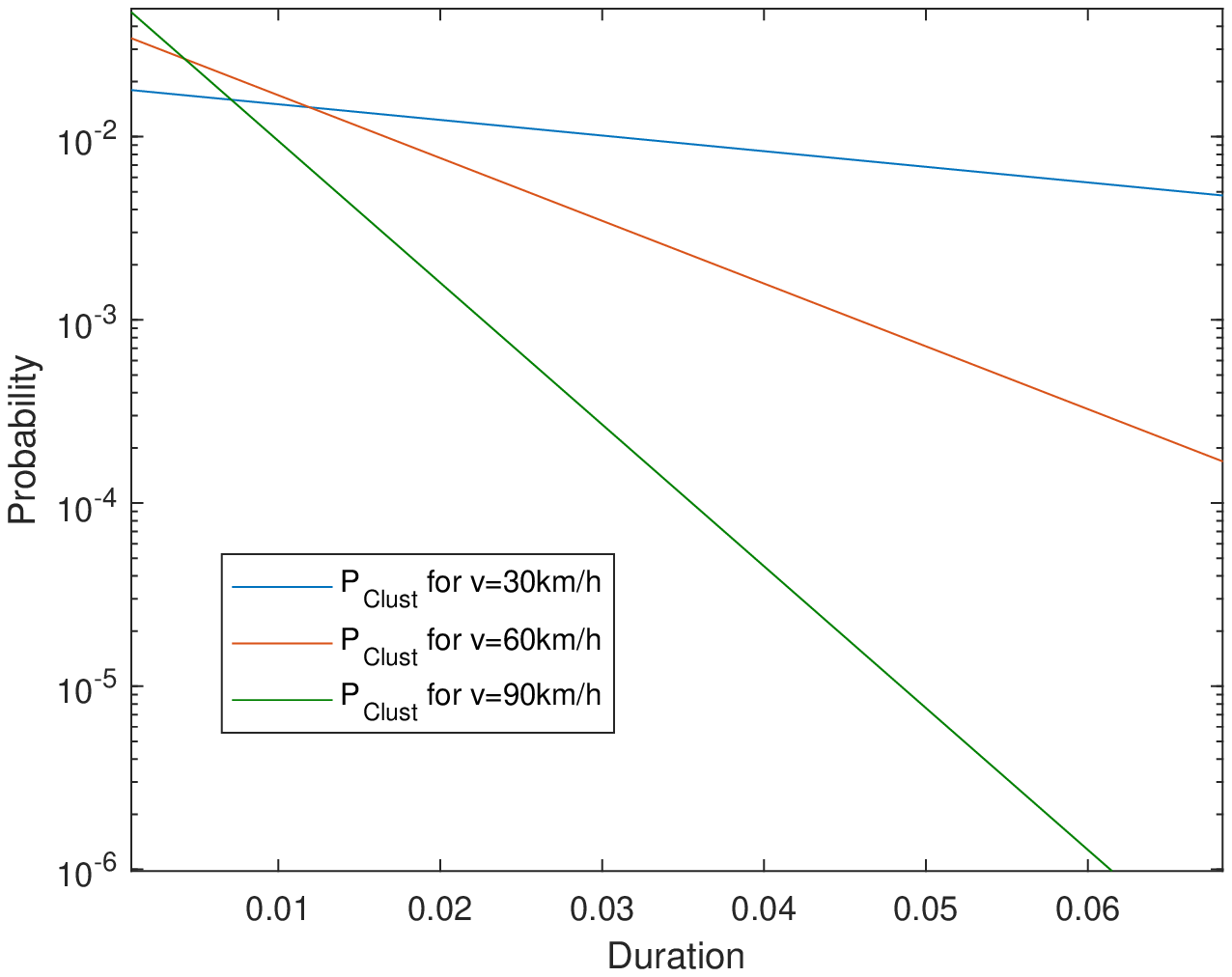}\\
	{\footnotesize Fig 4. Graphs of the probability $P_{clust}(m)$ given by the formula (\ref{eeeii}) for $v=30,60,90$ km/h}
\end{center}

\begin{center} %new2_checked2
	\includegraphics[scale=0.54]{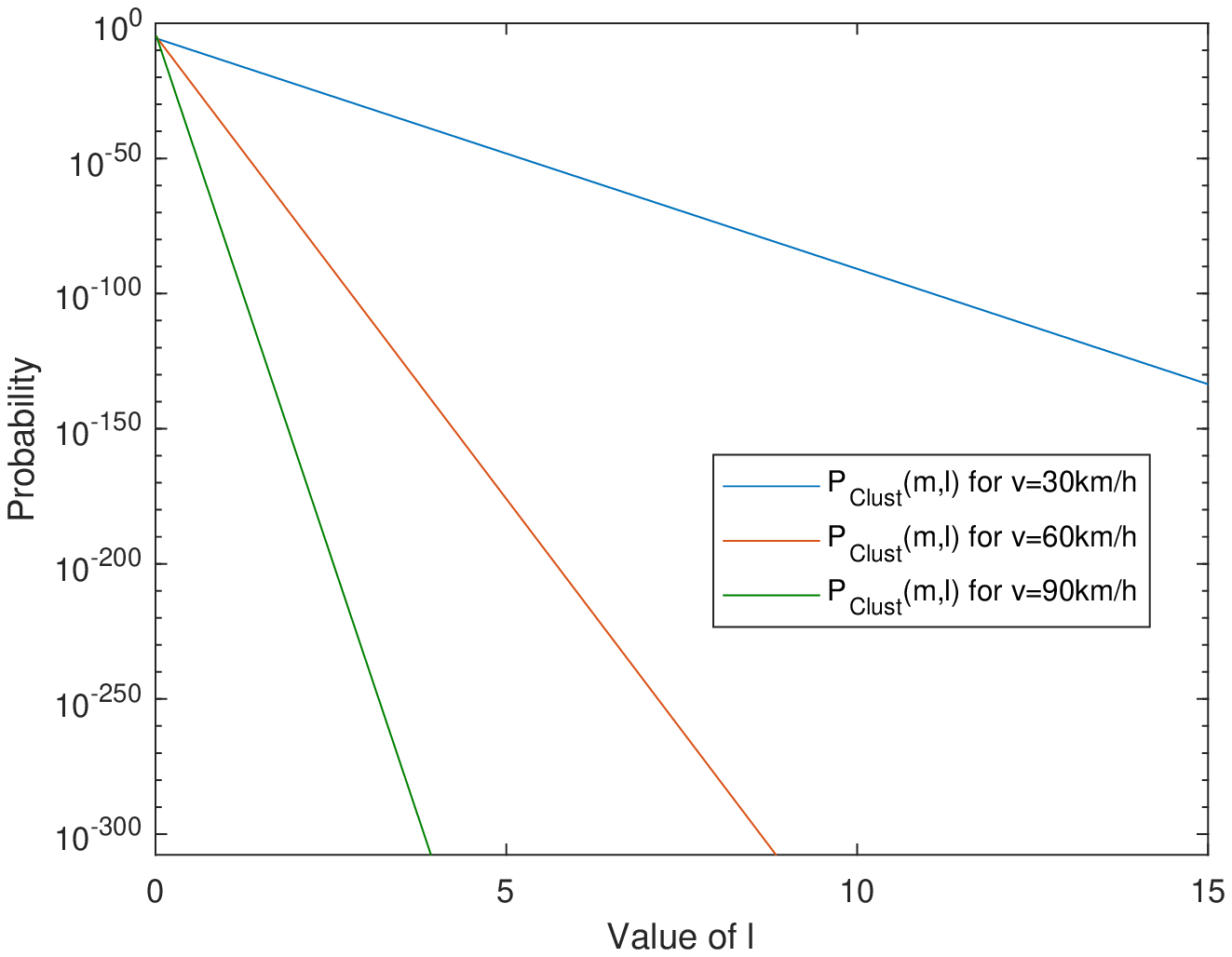}\\
	{\footnotesize Fig 5. Graphs of the probability $P_{clust}(r,t)$ of the cluster existence between times given by the formula (\ref{444}).}
\end{center}
\begin{center} %log
	\includegraphics[scale=0.54]{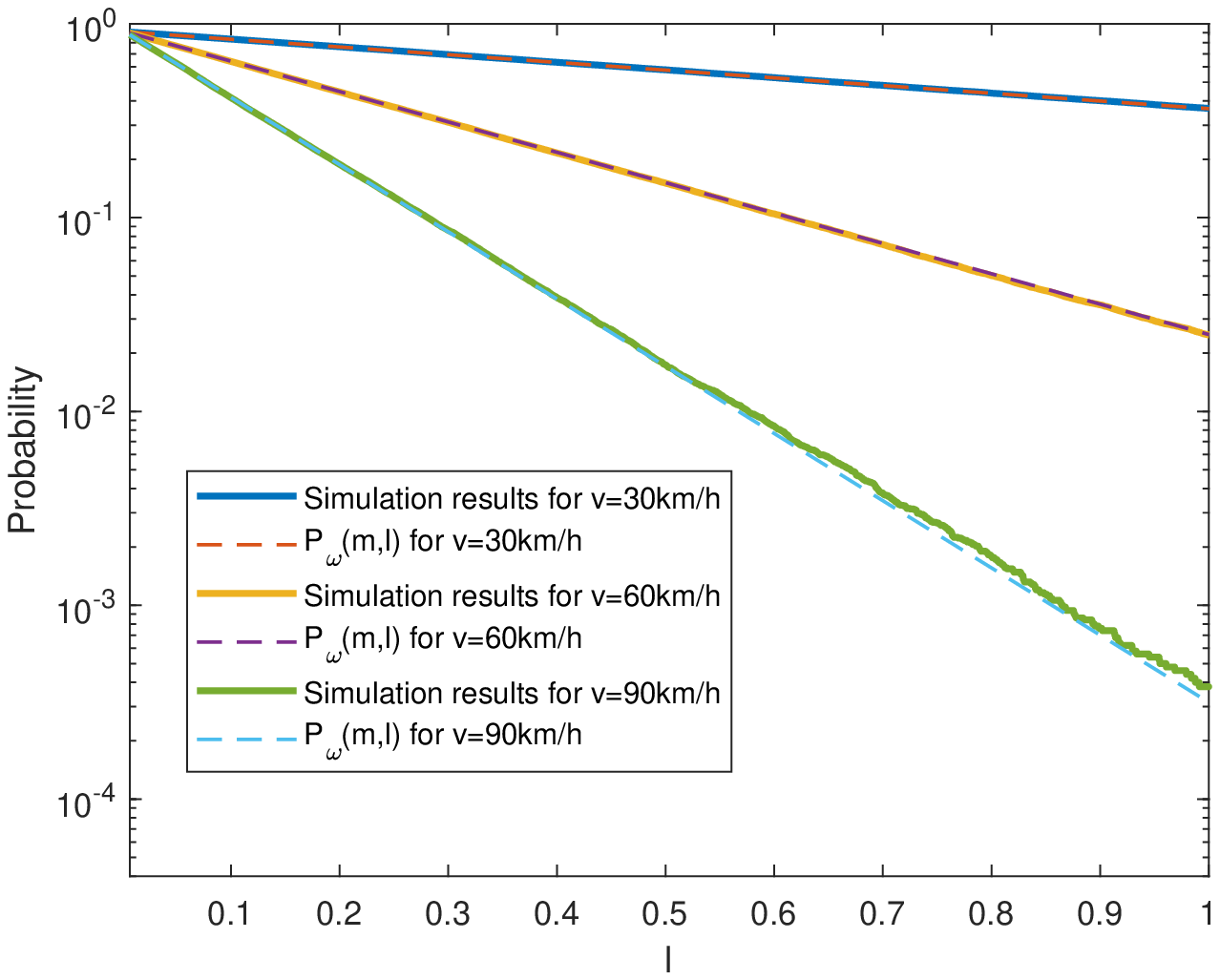}\\
	{\footnotesize Fig 6. Graphs of numerical simulation of the probability of 3-stable connection between times $m=2$ and $l$ with variable $l$, and the probability $P_\omega(m,l)$ returned by algorithm from the section \ref{Sec:StCon}.}
\end{center}

\section{Simulations for large values \\of parameters $p$ and $q$} \label{Sec:Sim2}

%Figs 3, 4 and 5 above do not include simulation results due the fact that they are not possible to perform in reasonable time due to negligible value of calculating probabilities (more detailed explanation see in the previous section).

In the section, we perform simulations that confirm the correctness of the formulas (\ref{eee}), (\ref{eeeii}) and (\ref{444}) for increased values of parameters $p$ and $q$, which have the order of $10^{-2}$. The results are represented on Figs. 7, 8 and 9 below.
The graphs perfectly match each other, which proves the correctness of the above mentioned formulas.

%The figures below show comparison between simulation results and probabilities predicted by the formulas (\ref{eee}), %(\ref{eeeii}) and (\ref{444}).  As it was mentioned before, the

Fig. 7 presents two graphs of the numerical simulation and probability of the link duration $P_{twocars}$ given by the formula (\ref{eee}) for $p=0.02$, $q=0.02$, $\Delta t=0.01$ and $10^5$ iterations.

Below on Fig. 8, the simulation results of the cluster lifetime duration are presented for the cluster that consists of cars $2$, $3$ and $4$, and also, the graph of the predicted distribution  $P_{clust}(m)$ for $p=0.02$,$q=0.02$, $\Delta t=0.01$ and $10^5$ iterations.

\begin{center}%new2_checked2
	\includegraphics[scale=0.54]{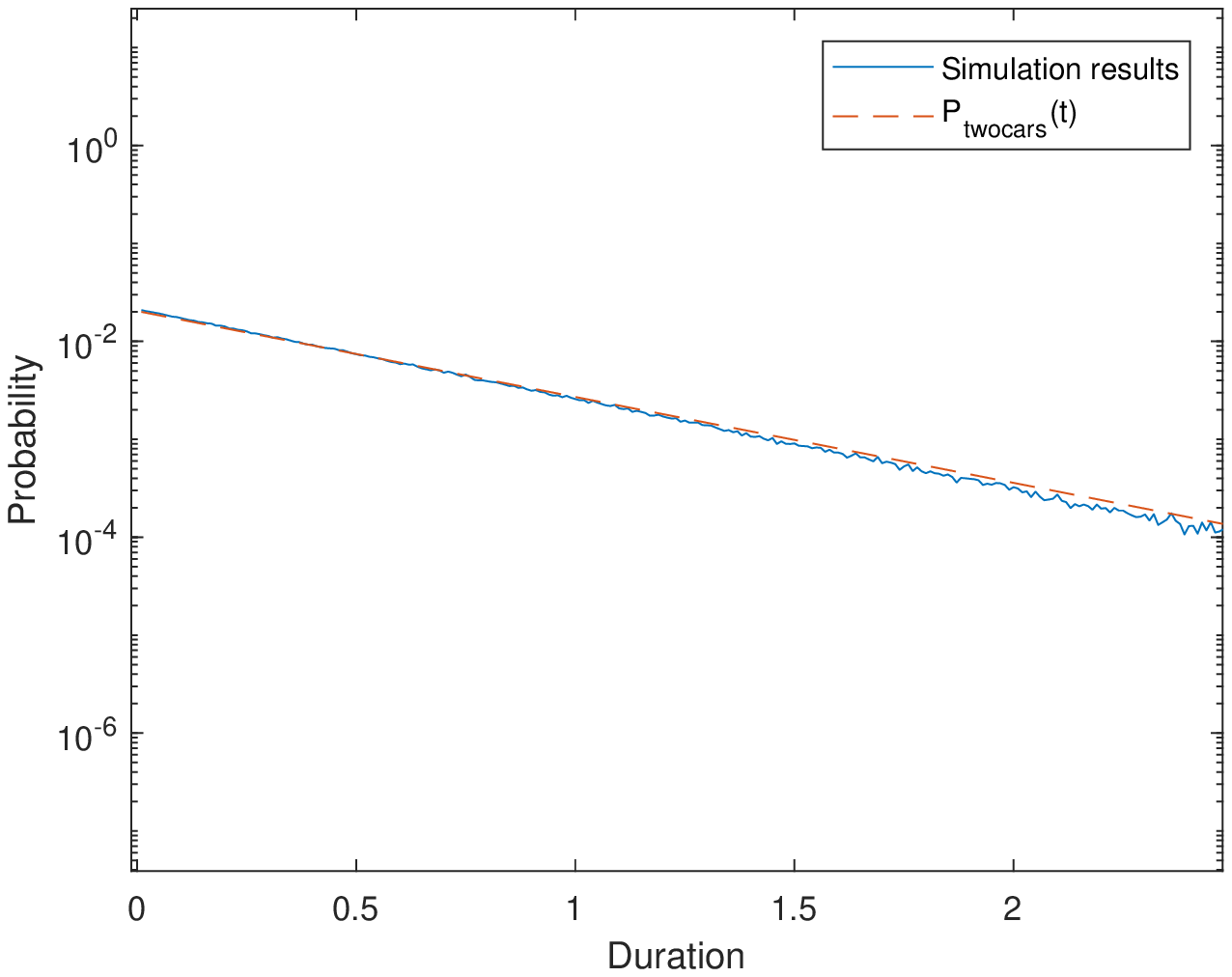}\\
	{\footnotesize Fig 7. Graphs of the numerical simulation of the distribution of the connection lifetime duration between two consecutive cars and distribution (\ref{eee}).}
\end{center}

\begin{center}%new2_checked2
	\includegraphics[scale=0.54]{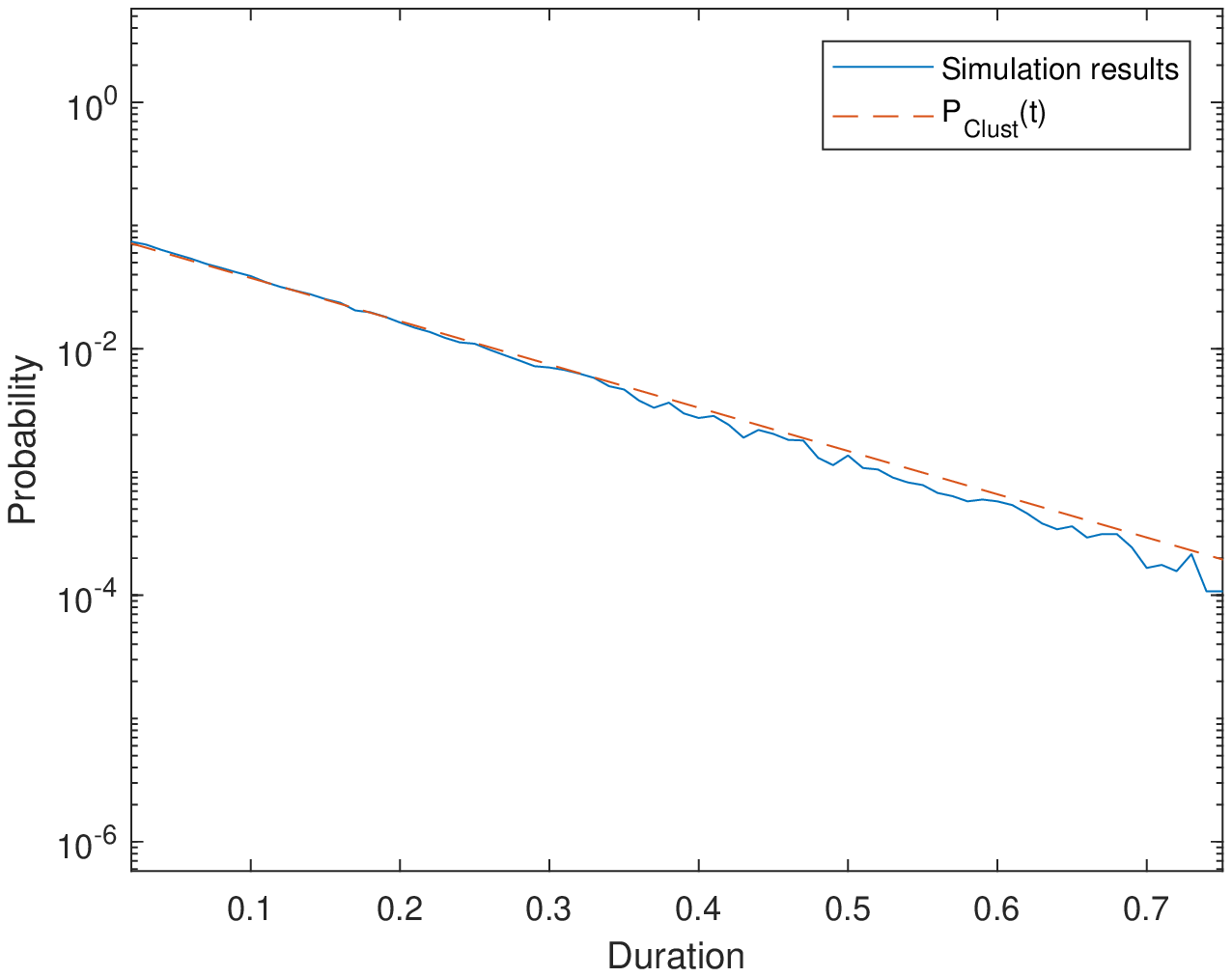}\\
	{\footnotesize Fig 8. Graphs of the numerical simulations of the cluster lifetime and the probability $P_{clust}(m)$.}
\end{center}

Simulation results of the probability of the cluster existence between times $15\Delta t$ and $15\Delta t,16\Delta t\ldots 30\Delta t$ are shown on Fig 9.
Thus, we assume value of $m=15$ to be constant and change the value of the parameter $l=15,16,\ldots 30$.
Fig. 9 depicts the graph of the probability (\ref{444}) for $p=0.05$, $q=0.05$, $\Delta t=0.01$ and $10^6$ iterations.

\begin{center}%log
	\includegraphics[scale=0.54]{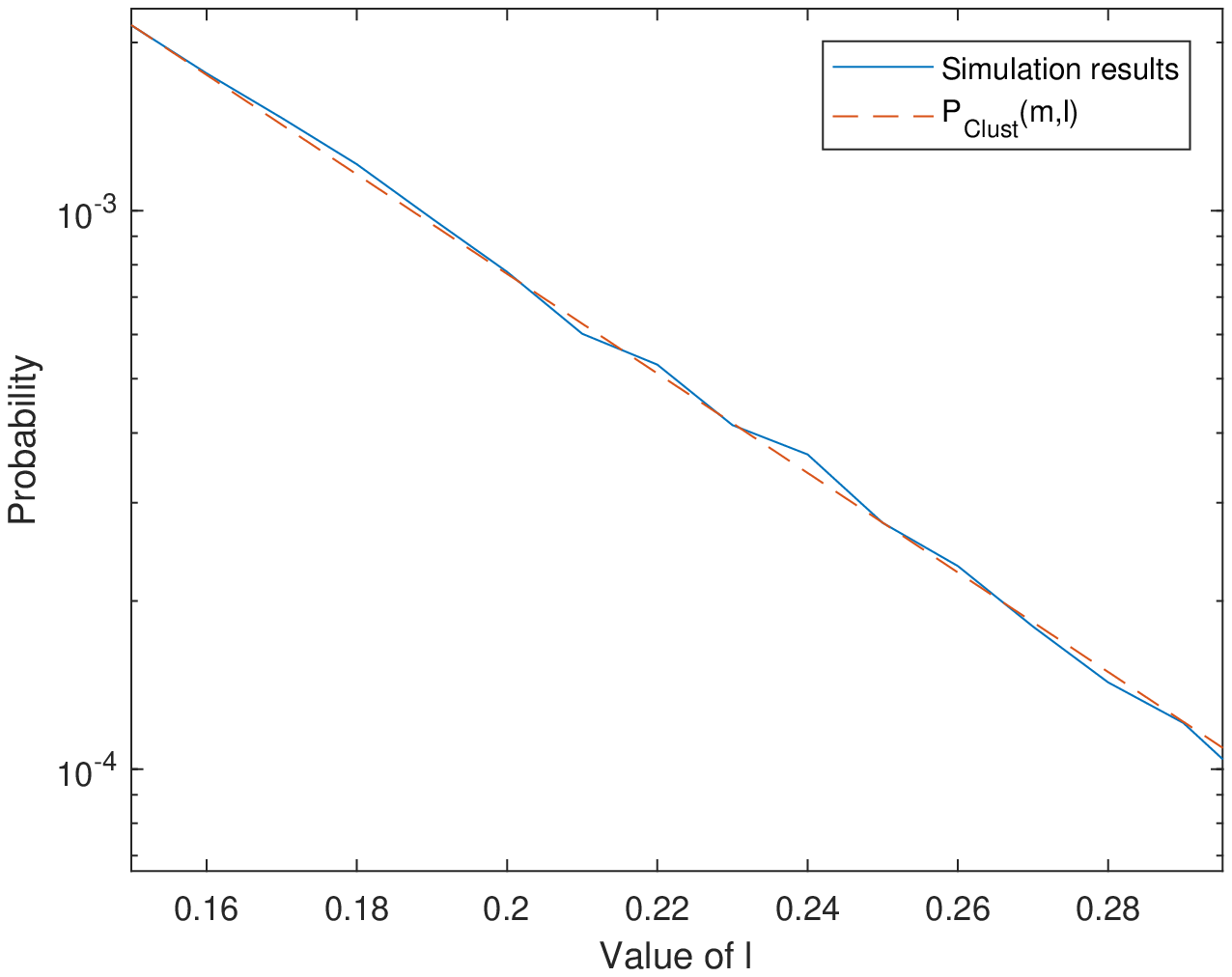}\\
	{\footnotesize Fig 9. Graphs of the numerical simulation of the probability of the cluster existence between times $15\Delta t$ and $15\Delta t,16\Delta t\ldots 30\Delta t$ and $P_{clust}(m,l)$.}
\end{center}

\section{Conclusion}

In the article, we consider evolution of the vehicle network on a highway and assume that the connection between each pair of consecutive cars can be established with a certain probability.
We derive the probability distributions that describe evolution of such characteristics of the network as a distribution of the link duration between each pair of cars, cluster lifetime, and \textit{etc}.
All the derivations are performed under the assumption that the connectivity model is described by the two state Markov chain model, where the parameters of the model are explicitly expressed through the parameters of the network.

\ifCLASSOPTIONcaptionsoff
  \newpage
\fi

\begin{IEEEbiography}{Gleb Dubosarskii}
received his B.E. and M.S. degrees in Applied Mathematics from the Ural Federal University in 2009 and 2011,
respectively, and his Candidate of Science Degree in Approximation Theory
from the Krasovskii Institute of Mathematics and Mechanics of the Ural
Branch of the Russian Academy of Sciences in 2014. Currently, he is a
Ph.D. student in the Department of Electrical and Computer Engineering
at Western University, Canada. His current research interests are in
the areas of communications, machine learning and mathematical
modeling.

\end{IEEEbiography}

\begin{IEEEbiography}{Serguei L. Primak}
(S’94-M’97) received the MSEE degree from
St. Petersburg University of Telecommunications, St. Petersburg,
Russia, in 1991 and the PhD degree in electrical engineering from the
Ben-Gurion University of the Negev, Beer-Sheva, Israel, in 1996.
Currently, he is a lecturer and post-doctoral fellow with the University of
Western Ontario, London, Ontario, Canada. His current interests are in
the field of ultrawideband radar applications, random signal generations,
modeling of wave propagation in a city, timefrequency analysis, and
inverse problems of electromagnetics. He is a member of the IEEE.
\end{IEEEbiography}

\begin{IEEEbiography}{Xianbin Wang}(S’98-M’99-SM’06-F’17) is a Professor and Tier 1 Canada Research Chair at Western
University, Canada. He received his Ph.D. degree in
electrical and computer engineering from National
University of Singapore in 2001.
Prior to joining Western, he was with Communications Research Centre Canada (CRC) as a Research Scientist/Senior Research Scientist between
July 2002 and Dec. 2007. From Jan. 2001 to July
2002, he was a system designer at STMicroelectronics. His current research interests include 5G
technologies, Internet-of-Things, communications security, machine learning
and locationing technologies. Dr. Wang has over 350 peer-reviewed journal
and conference papers, in addition to 29 granted and pending patents and
several standard contributions.
Dr. Wang is a Fellow of Canadian Academy of Engineering, a Fellow of
IEEE and an IEEE Distinguished Lecturer. He has received many awards and
recognitions, including Canada Research Chair, CRC Presidents Excellence
Award, Canadian Federal Government Public Service Award, Ontario Early
Researcher Award and six IEEE Best Paper Awards. He currently serves
as an Editor/Associate Editor for IEEE Transactions on Communications,
IEEE Transactions on Broadcasting, and IEEE Transactions on Vehicular
Technology and He was also an Associate Editor for IEEE Transactions
on Wireless Communications between 2007 and 2011, and IEEE Wireless
Communications Letters between 2011 and 2016. Dr. Wang was involved in
many IEEE conferences including GLOBECOM, ICC, VTC, PIMRC, WCNC
and CWIT, in different roles such as symposium chair, tutorial instructor, track
chair, session chair and TPC co-chair.
\end{IEEEbiography}
%It is not necessary to upload the biography when you submit your manuscript.

\end{document}